\documentclass[12 pt]{article}

\usepackage[dvips]{graphicx}
\usepackage{amssymb}
\usepackage{feynmf}

\def\fo{\hbox{{1}\kern-.25em\hbox{l}}}

\def\beq{\begin{equation}} \def\eeq{\end{equation}}
\def\eq{\end{equation}}

\newcommand{\gsim}{\lower.7ex\hbox{$\;\stackrel{\textstyle>}{\sim}\;$}}
\newcommand{\lsim}{\lower.7ex\hbox{$\;\stackrel{\textstyle<}{\sim}\;$}}

\def\beq{\begin{equation}} \def\eeq{\end{equation}}

\def\slashchar#1{\setbox0=\hbox{$#1$}           
   \dimen0=\wd0                                 
   \setbox1=\hbox{/} \dimen1=\wd1               
   \ifdim\dimen0>\dimen1                        
      \rlap{\hbox to \dimen0{\hfil/\hfil}}      
      #1                                        
   \else                                        
      \rlap{\hbox to \dimen1{\hfil$#1$\hfil}}   
      /                                         
   \fi}                                         %
\catcode`@=11
\long\def\@caption#1[#2]#3{\par\addcontentsline{\csname
  ext@#1\endcsname}{#1}{\protect\numberline{\csname
  the#1\endcsname}{\ignorespaces #2}}\begingroup
    \small
    \@parboxrestore
    \@makecaption{\csname fnum@#1\endcsname}{\ignorespaces #3}\par
  \endgroup}

\begin{document}

\begin{titlepage}

\begin{flushright}
UCD-2001-07\\
\end{flushright}

\huge
\begin{center}
{\Large\bf The impact of lepton-flavor violating $Z'$ bosons on muon $g-2$
and other muon observables}
\end{center}

\large

\vspace{.15in}
\begin{center}

Brandon Murakami

\small

\vspace{.1in}
\emph{Davis Institute for High Energy Physics, Department of Physics\\
University of California, One Shields Avenue, Davis, California 95616}
\end{center}
 
\vspace{0.15in}
 
\begin{abstract}
A lepton-flavor violating (LFV) $Z'$ boson may mimic some of the phenomena
usually attributed to supersymmetric theories.  Using a conservative model
of LFV $Z'$ bosons, the recent BNL E821 muon $g-2$ deviation allows
for a LFV $Z'$ interpretation with a boson mass up to 4.8 TeV while
staying within limits set by muon conversion, $\mu \to e\gamma$, and $\mu
\to eee$.  This model is immediately testable as one to twenty $e^+e^-
\to \mu\tau$ events are predicted for an analysis of the LEP II data.
Future muon conversion experiments, MECO and PRIME, are
demonstrated to have potential to probe very high boson masses with
very small charges, such as a 10 TeV boson with an $e$-$\mu$ charge of
$10^{-5}$. Furthermore, the next linear collider is shown to be highly
complementary with muon conversion experiments, which are shown to provide the
strictest and most relevant bounds on LFV phenomena.
\end{abstract}

\begin{flushleft}
hep-ph/0110095 \\
October 2001
\end{flushleft}

\end{titlepage}

\baselineskip=18pt

\newcommand{\lsp}{{\tilde\chi}} \newcommand{\squark}{{\tilde q}}
\newcommand{\lag}{{\begin{mathcal} L \end{mathcal}}}

\section{Motivation}

When surveying the possibilities for a fundamental theory of nature,
extra $U(1)$ gauge symmetries and their quanta, generically dubbed
$Z'$ bosons, are often found to naturally exist in the best of the current
extensions of the standard model (SM).  A renewed interest is urged by
recent and future experiments, including the BNL E821 muon anomalous
magnetic moment $2.6\sigma$ deviation \cite{Brown:2001mg}; the
forthcoming muon conversion experiments, MECO and PRIME \cite{meco,
prime}; the Tevatron and LHC hadronic colliders; and an anticipated
next linear collider.  Theoretically, we know of the following sources
for $Z'$ bosons:

\begin{itemize}
\item
The implied crossing of the three standard model gauge couplings
strongly hint at gauge coupling unification at some high energy scale.
Theories including such unification (GUTs) will, in general, include
non-SM $Z'$ gauge bosons.

\item
Theories with compactified extra spatial dimensions
\cite{Arkani-Hamed:1998rs, Randall:1999ee} provide many theoretical
tools:  a new interpretation of the gauge hierarchy problem, electroweak
symmetry breaking alternative mechanisms \cite{Dobrescu:1999dg}, and
supersymmetry breaking alternative mechanisms
\cite{Delgado:1999qr}.  The photon, SM $Z$ boson, or other neutral
vector bosons may form standing waves in the compactified extra
dimension.  While these excited states may or may not be derived
from new gauge symmetries, they will qualitatively mimic $Z'$
phenomenology.

\item
Supersymmetry is a highly motivated extension of the SM
\cite{Haber:1985rc, Martin:1997ns}.
Amongst a long list, a supersymmetric SM stabilizes the hierarchy
problem, explains electroweak symmetry breaking, naturally provides a
cold dark matter candidate, and provides a structure that permits
baron asymmetry solutions.  Particle multiplets in supersymmetric SM
theories with $N>1$ will include additional vector bosons.

\item
String theory is a candidate for a quantum theory of
gravity.  Supersymmetry and compactified extra dimensions may
naturally arise in some string models.  If the SM is imbedded within a
string model, large gauge groups, such as $E_8 \times E_8$ or
$SO(32)$, are often required.

\item
Strongly coupled theories  provide a dynamical electroweak symmetry
breaking mechanism, eliminate the need for scalar fields (considered
unnatural by some), and stabilize the weak scale from renormalizing
to the Plank mass \cite{Lane:2000pa}.  Through the extended gauge
structure at the heart of strongly coupled theories, extra $Z'$ gauge
bosons may emerge.
\end{itemize}

Historically, flavor-conserving $Z'$ bosons have been extensively
studied, while less attention has been given to the more general
case of $Z'$ bosons as a primal source of flavor changing neutral
currents (FCNC).  With proper respect for the discovery potential of the
forthcoming muon conversion experiments, we focus this article
only on the $Z'$ boson with lepton-flavor violation (LFV) at its
primitive vertices.  It is feasible that any of the above higher
theories will yield a LFV $Z'$ through sufficient model building and
interest.  In some cases, they already exist in the string models of
Refs.\,\cite{Nardi:1993nq, Cleaver:2000xe, Chaudhuri:1995cd},
technicolor models of Refs.\,\cite{Rador:1999is, Yue:2000fh}, and the
topflavor model \cite{Muller:1996dj}.
Because there is no general reason why the grander theories above
should supply flavor conserving $Z'$ bosons, the plausibility of LFV
$Z'$ bosons is inherited from the motivations of such models.

Further motivation stems from the strong possibility of detection of
weak-scale supersymmetric particles at near-future collider
experiments.  A supersymmetric analysis of collider signals may be
drastically altered by low-scale $Z'$ bosons by providing additional
process channels.  We demonstrate examples of this by showing how a
LFV $Z'$ boson may fully account for any deviations found in lepton
anomalous magnetic moments, muon conversion, $\mu \to e\gamma$, and
$\mu \to eee$.

\section{Choice of model}

The most general model of an electrically neutral $Z'$ bosons would
include in its lagrangian
its 1) kinetic term; 2) fermion interactions
$\frac{g_{Z'}}{\sin\theta_W}[\bar\psi_i \gamma^\mu (P_L
Q^{\psi_L}_{ij} + P_R Q^{\psi_R}_{ij}) \psi'_j] Z'_\mu$; 3) a
Higgs sector as a source for the $Z'$ mass; 4) a non-SM fermion sector
necessary to cancel the chiral anomalies of the $Z'$; 5) vector boson
interactions, including mass-mixing with the
$Z$ and other $U(1)$ bosons; and 6) kinetic mixing terms with other
vector bosons (i.e. $\frac{\chi}{2}Z^{\mu\nu}Z'_{\mu\nu}$).  The coupling
constant is chosen to have $\sin\theta_W$ separated out to make
comparisons to the SM $Z$ boson easier.  $\psi$ and $\psi'$ are labels for
leptons, up-type quarks, and down-type quarks:  $\psi \in \{l,u,d\}$.
The subscript on $\psi_i$ refers to flavor.  These charges must form
symmetric matrices if the lagrangian is to be Hermitian.

We choose a conservative model, meaning as few parameters
as possible while maintaining generic features that should be inherent
in any LFV $Z'$ boson.  We dub our choice of parameters ``model
$X$'', and the $Z'$ boson shares the same name, $X^\mu$.  To be
minimal, effects from the new Higgs and fermion sectors are considered
negligible, though their phenomena may be very rich in general.  The
$\chi$ parameter of gauge kinetic mixing is shown to be on the orders
of $|\chi|^2=10^{-6}$ for gauge-mediated supersymmetry breaking and
$10^{-16}$ for gravity-mediated supersymmetry breaking
\cite{Dienes:1997zr}.  Relative to
the dominant diagrams for any process considered here,
inclusion of kinetic mixing would suppress the contribution by
$|\chi|^2$.  There is no kinetic mixing in model $X$; we set $\chi=0$.
Another source of mixing would exist if the SM $Z$ boson
and $X$ boson shared a common Higgs mechanism.  The physical states
observed would then be a quantum admixture, parameterized by a mixing
angle $\theta$.  The physical $Z$ boson would then inherit some LFV
couplings.  No mixing is used in model $X$; $\theta=0$. 

We choose purely vectorial interactions for all fermions with the $X$
boson and drop the helicity subscripts on the charges while setting
$Q^{\psi_L}=Q^{\psi_R}$.  Separate charges for left- and right-handed
vector interactions offer no potential to change phenomenology since
spin-averaged observables will contain $|Q^{\psi_L}_{ij}|^2 +
|Q^{\psi_R}_{ij}|^2$ which is better off mapped to a single vectorial
coupling.  A similar mapping can be made for other choices,
i.e. purely axial, purely left-handed, etc.  Though, in purely
left- or right-handed interactions, processes that require a helicity
flip (i.e. mass insertions) will result in zero for an observable and
may be useful for restricting the classes of experiments that may
constrain a model, as in Ref. \cite{ourselves}.  For non-spin
averaged observables, such as muon $g-2$, more general coupling
choices may slightly alter phenomenology; Ref.~\cite{Choudhury:2001ad}
found muon $g-2$ compatibility at $2\sigma$ for the case of purely
vectorial couplings, but no compatibility at $2\sigma$ for the case of
purely axial couplings.

As our focus is LFV, we remove FCNC quark interactions by setting the
quark charge matrices to identity, $Q^u=Q^d=1$.
It assumed the higher theories that provide our $X$ boson presents all
fermions in the gauge basis of $X$---in general, different from the
basis of the SM (ordinary quarks and leptons) with
the Yukawa couplings in a non-diagonal basis.  The higher theory
supplies the $X$ charges in diagonal matrices $q^\psi$ (helicity label
omitted).  Through unitary matrices $U^{\psi}$, the original fermions
$\psi'$ are rotated to the SM basis $\psi=U^\psi\psi'$.  As a result,
the $X$ charges are transformed from $q^\psi$ to $Q^\psi =
U^{\psi\dagger} q U^\psi$.  The notation for fermion-vector boson
interactions is

\begin{equation}
{\cal L} \supset
\frac{g_X}{\sin\theta_W}
[\bar\psi'_i q^{\psi'}_{ij} \gamma^\mu \psi'_j] X_\mu
=
\frac{g_X}{\sin\theta_W}
[\bar\psi_i Q^\psi_{ij} \gamma^\mu \psi_j] X_\mu.
\end{equation}

\noindent
We will refer to the $Q_{ij}$ as ``charges'' even though they are not
the eigenvalues of a gauge group generator, but rather a simple
renaming of the quantity $[U^\dagger qU]_{ij}$.

The charges for the leptons $Q^l_{ij}$ contain the LFV content.
For the lepton charges, $q^l$ is chosen to be the
diagonal matrix $q^l={\rm diag}(q^l_{11}, q^l_{11}, q^l_{33})$ with
the first two generations sharing the same charge but different from
the third generation.  This is chosen for two reasons:  from the
fermion masses, one suspects there to be something special about the
third generation; and allowing the first two generations to have
unique charges does not affect generic behavior in any qualitative way.

All Yukawa unitary rotation matrices are set to the identity matrix
except the following two.   $U^{u_L} = V_{\rm CKM}$ is necessary to
meet the definition of the CKM matrix.  For the lepton charges,
a non-trivial $U^{l}$ will
create off diagonal elements from the diagonal $q^l$ matrix provided
$q^l$ is not proportional to the identity matrix.  $U^l$ is chosen to
borrow the parameterized form of the CKM matrix,

\beq U^{l} = \left( \begin{array}{ccc} c_{12}c_{13} & s_{12}c_{13} &
s_{13} \\ -s_{12}c_{23}-c_{12}s_{23}s_{13} &
c_{12}c_{23}-s_{12}s_{23}s_{13} & s_{23}c_{13} \\
s_{12}s_{23}-c_{12}c_{23}s_{13} & c_{12}s_{23}-s_{12}c_{23}s_{13} &
c_{23}c_{13}
\end{array} \right).
\eeq

\noindent
The notation $s_{ij}$ and $c_{ij}$ means sines and cosines of
parameters $\theta_{12}$, $\theta_{23}$, $\theta_{13}$ which need not
be the CKM values.  We have ignored the allowed complex phase for
simplicity.  In summary, model $X$ is defined by parameters $q^l_{11}$,
$q^l_{33}$, three angles for the unitary transformation $U^l$, the
coupling constant $g_X$, and the boson mass $m_X$.

\section{Experimental constraints on LFV}

Although it is known that LFV may occur in other
models such as slepton loops in a supersymmetric SM, loops with
neutrino mixing, and other exotic models \cite{Kuno:2001jp}, it
assumed that the $X$ boson is the dominant component of LFV effects in
the following analysis.  With neutrino mixing recently verified to
3$\sigma$ for active and sterile neutrino models \cite{Bahcall:2001zu,
Barger:2001zs}, LFV
is a reality.  But to what extent?  It can be shown that LFV through
neutrino mixing cannot account for any LFV signals found with the
sensitivity levels of the future muon conversion experiments.
However, a supersymmetric SM enhances LFV effects of neutrino
mixing in see-saw mechanisms through Yukawa vertices of the type
$H^\pm-l^\mp-\nu$ \cite{Hisano:1999qm}.

\begin{table}
\center
\begin{tabular}{|c|c|c|}
\hline
Experiment type & Charges probed & Best measurement\\
\hline\hline
muon $g-2$ & $Q^l_{23}$, ($Q^l_{21}$, $Q^l_{22}$) & $\delta a_\mu =
(43\pm16)\times10^{-10}$\\
\hline
$\mu N \to eN$ & $Q^l_{12}$ & $R<6.1\times10^{-13}$ \\
\hline
$\mu \to eee$ & $Q^l_{11}$, $Q^l_{12}$ & $R<1.0\times10^{-12}$  \\
\hline
$\mu \to e\gamma$ & $Q^l_{13}$, $Q^l_{23}$, (and all others) &
$R<1.2\times10^{-11}$\\
\hline
$e^+e^- \to \mu\tau$ & $Q^l_{11}$, $Q^l_{23}$, ($Q^l_{12}$,
$Q^l_{13}$) & n/a \\
\hline
$e^+e^- \to \mu^+\mu^-$ & $Q^l_{11}$, $Q^l_{22}$, ($Q^l_{12}$) & n/a\\
\hline
\end{tabular}
\caption{The charges listed are only those involved at lowest order.
Those in parenthesis are involved in diagrams of the same order, but
suppressed by either a mass insertion or the charge expected to be
small.  The quantity $R$ is defined as  $\sigma(\mu N \to e
N)/\sigma(\mu N \to \nu_\mu N)$ for muon conversion and as the branching
ratios for the rare muon decays.}
\end{table}

Table 1 lists the experiments considered and the charges they probe at
lowest perturbative order.  Note that, in general, they all probe
different charges.  Since all measurements have null-results except
for the muon anomalous magnetic moment, we may consider two cases: 
the case in which the $X$ boson contributes to the muon anomalous
magnetic moment, but does not dominate; and the case in which it
does dominate.  The reason for this division is due to the great
difference in available parameter space between the two cases---the
latter case being highly constrained.

In general, for model $X$, a single experiment from Table 1 has its
charges constrained by any other experiment listed in Table 1.  For
example, $\mu \to eee$ shares the at least one charge in common as
$e^+e^- \to \mu\tau$ and muon conversion; $e^+e^- \to \mu\tau$ shares
at least one charge in common with a few other experiments, and so on
until all six charges are included.  There are exceptions to this general
feature of model $X$.  This arises when a parameter choice sets one or
more charges $Q^l_{ij}$ to zero, such as the universal or
generation-dependent $Z'$ cases of model $X$.  For example, setting
$Q^l_{12}=0$ unconstrains a model X interpretation of muon $g-2$ from
the strict muon conversion and $\mu\to e\gamma$ limits.

\subsection{The muon anomalous magnetic moment}

\begin{figure}
\center
\begin{fmffile}{g2}
\begin{fmfgraph*}(200,75)
\fmfleft{muin}
\fmfright{muout}
\fmfbottom{ext}
\fmf{fermion,label=$\mu^+$}{muout,v1}
\fmf{fermion,label=$l^+_i$}{v1,v2}
\fmf{fermion,label=$l^+_i$}{v2,v3}
\fmf{fermion,label=$\mu^+$}{v3,muin}
\fmffreeze
\fmf{photon,right,tension=0,label=$X$}{v1,v3}
\fmf{photon,label=$\gamma$}{v2,ext}
\fmfdot{v1,v2,v3}
\end{fmfgraph*}
\end{fmffile}

\caption{The leading contributions of a LFV $Z'$ boson to the muon
anomalous magnetic moment.  The diagram with an internal tau
propagator is dominant due to $m_\tau^2$ enhancement in the $g-2$
observable.}
\label{fig.g2}
\end{figure}
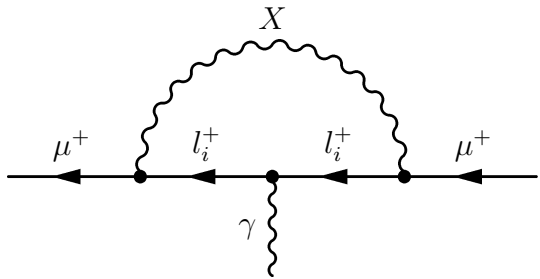

The $X$ boson participates in the (anti)muon anomalous magnetic moment via a
loop involving all charged leptons (Fig.\,\ref{fig.g2}).
The $X$ boson contribution is dominated by the diagram that includes
an internal tau line.  This can be seen by noting the two internal
tau propagators provide a term proportional to $m_\tau^2$ in the
numerator.  This dependence is not cancelled by the $m_\tau^2$ in the
denominator since a heavy $X$ boson mass will dominate the
denominator.  Therefore, we ignore
all other contributions by the $X$ boson and only the $\mu$-$\tau$
charge $Q^l_{23}$ is probed.  In the limit $m_X\gg m_\tau$, the
deviation of the muon anomalous magnetic moment $a_\mu
\equiv (g-2)/2$ from the SM value is

\begin{equation}
\delta a_\mu = \frac{g_X^2 (Q^l_{23})^2}{4\pi^2\sin^2\theta_W}\frac{m_\mu
m_\tau}{m_X^2}
\end{equation}

\noindent
and is measured to be $(43\pm16)\times10^{-10}$ by BNL experiment
E821 (2000).  Parameter space for
model $X$ to account for $\delta a_\mu$ exists but is small when
enforcing limits from the LFV experiments listed in Table 1.  This is
accomplished by large $\mu$-$\tau$ charge $Q^l_{23}$, small $e$-$\mu$
charge $Q^l_{12}$,
and relatively small $m_X$.  The boson mass $m_X$ must be balanced
between being light enough to account for $\delta a_\mu$ and heavy
enough to avoid being ruled out at muon conversion and $\mu \to eee$
experiments.  The small $Q^l_{12}$ criteria is required to keep
under the same limits.  To create a large $Q^l_{23}$, the special case
of $q^l={\rm diag}(-1,-1,1)$ is used since all off-diagonal charges are
proportional to the difference in $q^l_{11}$ and $q^l_{33}$ in model
$X$,

\begin{equation}
Q^l_{ij} = (q^l_{33}-q^l_{11}) U^{l\dagger}_{i3}U^l_{3j} \qquad (i\neq
j).
\end{equation}

Parameter space compatible with the BNL E821 deviation is tightly
bunched around a particular charge assignment

\begin{equation}
Q^l \approx \left(
\begin{array}{ccc}
-1 & {\cal O(}10^{-5}) & {\cal O}(10^{-5}) \\ & 0.1\,{\rm to}\,1 &
0.4\,{\rm to}\,1 \\ & & 0.1\,{\rm to}\,1
\end{array}
\right).
\end{equation}

\begin{figure}[!btp]
\center
\resizebox{4 in}{!}{\includegraphics{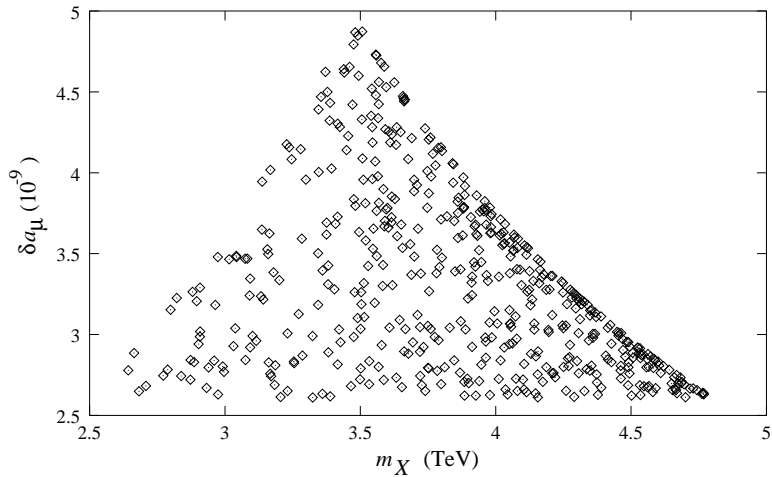}}
\caption{A scatter plot of the LFV boson mass $m_X$ (TeV) vs.~the muon
anomalous magnetic moment deviation $\delta a_\mu$.  Each point is a
point in the parameter space of a LFV $Z'$ boson theory that fully
accounts for the BNL E821 muon $g-2$ observed deviation.  All points
are within limits set by the LFV experiments of Table 1, with muon conversion
as the strictest limit.  ($\nu_\mu e \to \nu_\mu e$ elastic scattering
limits are not applied.)  The coupling constant is set to the maximal
value that retains the perturbative calculation's validity,
$g_X/\sin\theta_W=\sqrt{4\pi}$.  This is to demonstrate the highest
allowed boson mass.}
\label{fig.splot}
\end{figure}

\noindent
All entries above are magnitudes only.  However, a quirk of model $X$
requires the $Q^l_{11}$ entry to be fixed at $-1$ at its muon $g-2$
optimal compatible value, despite whether or not the case
$q^l=\textrm{diag}(-1,-1,1)$ is used.
This parameter space yields an upper limit on the boson mass of $m_X
\approx$1 TeV for $g_X = g_Y$ and $2.8$ TeV $\lesssim m_X \lesssim$4.8
TeV for $g_X/\sin\theta_W = \sqrt{4\pi}$, as seen in Fig.\,\ref{fig.splot}.
$\sqrt{4\pi}$ is chosen to be the highest value of the coupling
constant chosen such that next leading order in the perturbative expansion
remains smaller than the leading order contribution.  If the
constraint $q^l=\textrm{diag}(-1,-1,1)$ is relaxed, the effect is to
widen the range of allowed boson masses.

$Q^l_{11}$ fixed at $-1$ has potential to conflict with measurements of
elastic scattering of muon neutrinos and electrons, $\nu_\mu e \to
\nu_\mu e$. 
While not a charge involved in the lowest order $X$ contribution to
muon $g-2$, $Q^l_{11}$ is nonetheless constrained due to the
parameterization choice of model $X$.  To examine this constraint, the
4-point effective lagrangian is constructed from the $X$-exchange in
the $t$-channel,

\begin{equation}
\mathcal{L}_{\rm effective} \supset
\frac{g_X^2|Q^l_{11}|^2/\sin^2\theta_W}{t-m_X^2}
[\bar\nu_\mu \gamma^\lambda P_L \nu_\mu] [\bar e\gamma_\lambda e]
\end{equation}

\noindent
where $P_L=(1-\gamma^5)/2$.  This is done for the sake of comparison
with the convention of the PDG (Eq.~10.12 of Ref.~\cite{Groom:2000in}),

\begin{equation}
\mathcal{L}^{\nu e} = \frac{G_{\rm F}}{2}
[\bar\nu_\mu \gamma^\lambda (1-\gamma^5)\nu_\mu]
[\bar e \gamma_\lambda(g_V^{\nu e}-g_A^{\nu e}\gamma^5) e].
\end{equation}

\noindent
This constrains $|Q^l_{11}|$ to $g_V^{\nu e}$.  Using a typical
momentum transfer value $|t|\ll m_X^2$ and the PDG quoted value
of $g_V^{\nu e}=(-0.041 \pm 0.015)$, it is seen that there is no
model $X$ parameter space (namely $Q^l_{11}$) that simultaneously
satisfies the muon
$g-2$ deviation, muon LFV experiments, and $\nu_\mu e \to \nu_\mu e$
scattering measurements.  Therefore, in order for a LFV $Z'$
model to have compatibility with all such data, a modified model $X$ would
require either one more additional parameter to unconstrain $Q^l_{11}$
or arbitrarily small couplings to the muon neutrino.

Compatibility without $\nu_\mu e \to \nu_\mu e$ constraints is in
agreement with Ref. \cite{Huang:2001zx}, which uses a
submodel of model $X$ to demonstrate a LFV $Z'$ interpretation of the
muon $g-2$ deviation. However, the model of Ref. \cite{Huang:2001zx}
has the limitation of not being able to be tested at a linear collider
since it does not include electron couplings (i.e. $Q^l_{11}=0$).
Another recent study \cite{Lynch:2001zr} also utilized submodels of
model $X$ (non-commuting extended technicolor and top assisted
technicolor), but such models found only small LFV $Z'$ contributions
to the muon $g-2$ deviation.

In the case of diagonal charges in $Q^l$,  model $X$ becomes a
universal or generation-dependent $Z'$ study.  Such models still may
account for the muon $g-2$ deviation simply by having a large coupling
to the muon.  The internal fermion propagator is a muon, and the muon
$g-2$ deviation is then

\begin{equation}
\delta a_\mu = \frac{g_X^2
(Q^l_{22})^2}{12\pi^2\sin^2\theta_W}\frac{m_\mu^2}{m_X^2},
\end{equation}

\noindent
in the limit $m_X\gg m_\mu$.  Using $g_X/\sin^2\theta_W =
\sqrt{4\pi}$ and $|Q^l_{22}|=1$, a 660 GeV boson is allowed.  With a
more familiar value for the coupling constant, $g_X=g_Y$, the boson
mass would have to be under 140 GeV to explain the deviation.
Neither constraints from
LEP II and $\nu_\mu e \to \nu_\mu e$ (both probes of $Q^l_{11}$)
may not rule out these mass limits due to the possible
parameter space in which the electron charge $Q^l_{11}$ can be set
arbitrarily small while the muon charge $Q^l_{22}$ is arbitrary.  For
example, this is seen by the choice that $q^l={\rm diag}(0,0,q^l_{33})$ while
the unitary rotations are constrained to
$\theta_{12}=\theta_{13}=\pi/2$ with $\theta_{23}$ left arbitrary.
This results in $Q^l_{11}=0$ and arbitrary $Q^l_{22}$.

\subsection{$e^+e^- \to \mu\tau$}

With the LFV $X$ boson interpretation of the muon $g-2$ deviation
requiring a large off-diagonal charge $Q^l_{23}$, one immediately
wonders if model $X$ predicts $e^+e^- \to \mu\tau$ events observed at
LEP II. This cross-section is

\begin{equation}
\sigma(e^+e^- \to \mu\tau) =
\frac{g_X^4}{12\pi\sin^4\theta_W}(Q^l_{11}Q^l_{23})^2\frac{s}{(s-m_X)^2}.
\end{equation}

\noindent
Using a center of mass energy of 210 GeV and a total luminosity of 230
pb$^{-1}$, all parameter space points in the plots of
Fig.\,\ref{fig.splot} predict one to about twenty events at LEP II,
regardless of the coupling constant $g_X$.  This is fascinating and
suggests an analysis of the LEP II data would be elucidating.  If no
events are found at LEP II, model $X$ is still a plausible
interpretation for the muon $g-2$ deviation since an ample fraction of
the parameter space points yield only a handful of events. 

Considering $e^+e^- \to e\mu$ is pointless in the foreseeable future due
to the strict muon conversion limits.  $e^+e^- \to e\tau$ is
unmotivated in this study as there is no reason to believe the
$e$-$\tau$ charge $Q^l_{13}$
is large.  If a LFV $Z'$ is not involved in the dominant
contribution to the muon $g-2$ discrepancy, $Q^l_{13}$ may be large
but still unmotivated.

\subsection{Muon conversion}
Muon conversion is the process $\mu^- N \to e^- N$.  Slow negative
muons are aimed
at a nuclear sample where ground state muonic atoms are allowed to
form.  The muon eventually undergoes SM decay in which a $W$ boson is
emitted from the muon towards the nuclei or outside the atom.  The
ratio of muon conversion to weak decays is defined as $R(\mu N \to e
N) \equiv \sigma(\mu N \to e N)/\sigma(\mu N \to \nu_\mu N')$.
SINDRUM II at the Paul Scherrer Institut (PSI) holds the current best limit of
$6.1\times 10^{-13}$ (1998) \cite{sindrumii}.  MECO
(\underline{m}uon \underline{e}lectron \underline{co}nversion) at
Brookhaven (E940) may collect data in 2006 with a sensitivity of
$2\times 10^{-16}$ \cite{molzon}.  PRIME (\underline{PRI}SM
\underline{m}u-\underline{e} conversion) will use the PRISM
high-intensity muon source at the Japan Hadron Facility (to be
renamed) at KEK and may collect data in 2007 with a sensitivity of
$10^{-18}$ \cite{kuno}.  This great technological leap of more than 4
orders of magnitude warrants LFV as a larger part of our community's
consciousness for the next decade.

The muon conversion ratio is \cite{Langacker:2000ju}
\begin{eqnarray}
R(\mu N \to eN) &=& \frac{G_{\rm F}^2\alpha^3 m_\mu^5}{2\pi^2\Gamma_{\rm capture}}
\frac{Z_{\rm eff}^4}{Z}|F_P|^2 \left( |Q_{12}^{l_L}|^2
+|Q_{12}^{l_R}|^2 \right) \nonumber\\ && \times \left|
\frac{g_X}{g_Y}\sin\theta\cos\theta
\left(1-\frac{m_W^2}{m_X^2\cos^2\theta_W}\right) \left[
\frac{1}{2}(Z-N)-2Z\sin^2\theta_W \right] \right. \nonumber\\ &&
\left. + \frac{g_X^2}{g_Y^2} \left( \sin^2\theta +
\frac{m_W^2}{m_X^2\cos^2\theta_W} \cos^2\theta \right) \left[
(2Z+N)(|Q_{11}^{u_L}|^2 + |Q_{11}^{u_R}|^2) \right.\right. \nonumber\\
&&+ \left.\left. (Z+2N)(|Q_{11}^{d_L}|^2 + |Q_{11}^{d_R}|^2) \right]
\right|^2.
\end{eqnarray}

\noindent
Using $^{48}$Ti as the sample, nuclear form factor $F_P=0.54$
\cite{Bernabeu:1993ta}, $Z_{\rm eff}=17.6$ \cite{Sens}, and the muon
capture rate $\Gamma_{\rm capture}=2.6 \times 10^{-6}$ s$^{-1}$
\cite{Suzuki:1987jf}.  In model $X$, this simplifies to

\begin{equation}
R(\mu N \to e N) = 3.1\times10^{-11} \left( \frac{g_X}{g_Y} \right)^4
\left( \frac{Q^l_{12}}{10^{-5}} \right)^2 \left( \frac{1\,{\rm
TeV}}{m_X} \right)^4.
\end{equation}

\noindent
Note that $R \propto 1/m_X^4$ means for every four orders of
magnitude gained through the experimentalists' efforts, it becomes
possible to probe $Z'$ bosons exactly one magnitude more massive.
All processes that involve only one $Z'$ internal propagator will
share this feature in its observable, such as $\mu\to eee$ and $\mu\to
e\gamma$.

\begin{figure}[!btp]
\center \resizebox{4 in}{!}{\includegraphics{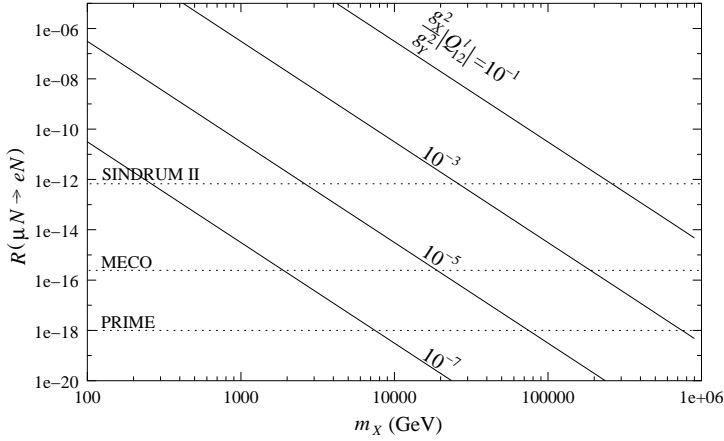}}
\caption{LFV boson mass $m_X$ (GeV) vs.\ the muon conversion rate
$R(\mu^-N \to e^-N)$. This plot demonstrates the discovery potential
of future muon conversion experiments.  The diagonal lines represent
different values of the $\mu e X$ vertex coupling.  Note the
surprisingly large boson mass and small couplings accessible by future
experiments MECO and PRIME.  This plot does not assume the LFV should
account for all of the BNL E821 muon $g-2$ observed deviation.}
\label{fig.muconv}
\end{figure}

The vast discovery potential of future muon conversion experiments is
demonstrated in Fig.\,\ref{fig.muconv}.  This plot and all plots to
follow do not assume that the model $X$ boson fully accounts for the
muon $g-2$ observed deviation.  Fig.\,\ref{fig.muconv} is best appreciated
by noting the large boson masses and small $e$-$\mu$ charges $Q^l_{12}$
accessible.  For example, a LFV signal at PRIME may imply a model $X$
boson with a mass about ${\cal O}(10\,{\rm TeV})$ and couplings as small
as ${\cal O}(10^{-5})$.  Even without a LFV signal, MECO and PRIME
will provide strict bounds on theoretical models that include LFV.

Due to the stringent current muon conversion limit, $Q^l_{12}$ is
constrained to be
very small for light $X$ bosons.  Because of the relationships between
all charges of model $X$, muon conversion constrains the parameter
space of model $X$ more so than any other experiment, as will be shown
in the following subsections.

\subsection{$\mu \to eee$}

This process has been historically performed using antimuons at rest,
$\mu^+ \to e^+e^+e^-$.  A negative muon will tend to be captured
by a nucleus in the target used to stop the muon, and is therefore not
used.  $\mu \to eee$ is similar to muon conversion in that they both
stringently probed the $e$-$\mu$ charge $Q^l_{12}$.  $\mu \to eee$
differs in that it also probes $e$-$e$ charge $Q^l_{11}$ at tree level.

The partial width for $\mu \to eee$ is

\begin{equation}
\Gamma(\mu \to eee) = \frac{G_{\rm F}^2 m_\mu^5}{4\pi^3} \left(
\frac{g_X}{g_Y} \right)^4 \left( Q^l_{11} Q^l_{12} \right)^2 \left(
\frac{m_Z}{m_X} \right)^4.
\end{equation}

\noindent
The branching ratio for model $X$ at tree level results in

\begin{equation}
R(\mu \to eee) = 3.4\times10^{-13} \left( \frac{g_X}{g_Y} \right)^4
\left( \frac{Q^l_{11} Q^l_{12}}{10^{-5}} \right)^2 \left(
\frac{1\,{\rm{TeV}}}{m_X} \right)^4.
\end{equation}

\noindent
The current sensitivity level is $R(\mu \to eee)<1.0\times10^{-12}$
from SINDRUM at PSI (1988) \cite{Bellgardt:1988du}.  There are no
major $\mu \to eee$ projects announced as forthcoming.  However, with
the future high-intensity muon sources in the works, a near-future
experiment is yet plausible.

\begin{figure}[!btp]
\center \resizebox{4 in}{!}{\includegraphics{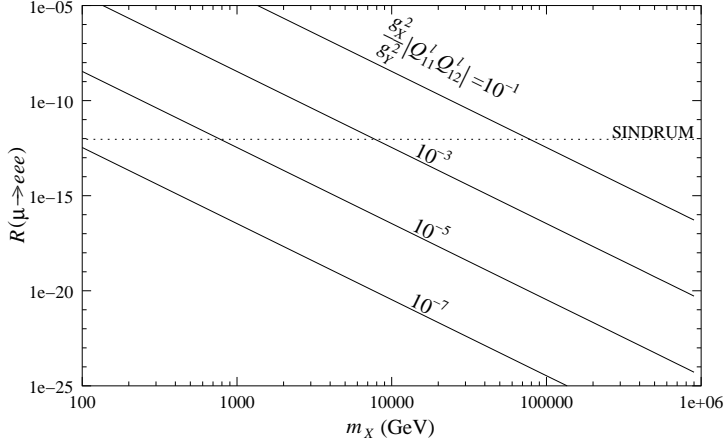}}
\caption{LFV boson mass $m_X$ (GeV) vs.\ the branching ratio $R(\mu
\to eee)$.  The value of this plot is understood if compared to
Fig. \ref{fig.muconv}.  It is obvious that the current limits of muon
conversion and $\mu \to eee$ are competitive in their parameter space
coverage.  More subtle is noting that, muon conversion has probed more
parameter space than $\mu \to eee$.  For example, using $Q^l_{11}=-1$ and
$Q^l_{12}=10^{-5}$ (the muon $g-2$ optimal parameters), muon
conversion has probed the LFV boson mass to 2 TeV while $\mu \to eee$
probed to 800 GeV.  Therefore, in the near future, muon conversion
warrants more attention as the MECO and PRIME experiments are coming
online in 2006 and 2007 respectively.  This plot does not
assume the LFV should account for all of the BNL E821 muon $g-2$
observed deviation.}
\label{fig.mu3e}
\end{figure}

Fig.\,\ref{fig.mu3e} shows us that
$\mu \to eee$ is a competitive experiment with muon conversion for
probing the $e$-$\mu$ charge $Q^l_{12}$.  Closer inspection
reveals muon conversion has probed more
parameter space than $\mu \to eee$.  For example, if the muon $g-2$
parameters of $Q^l_{11} = -1$ and $Q^l_{12} = 10^{-5}$ are used, $\mu
\to eee$ has probed $X$ bosons to 800 GeV while muon conversion has
probed to 2 TeV.  Furthermore, muon conversion has provided stricter
bounds than linear and hadronic collider searches for
generation-dependent $Z'$ bosons that originate from $SU(2)_h\times
SU(2)_l$ and $U(1)_l \times U(1)_h$ extended electroweak gauge
structures (special cases of model $X$), as studied in
Ref.\,\cite{Lynch:2001md, Chivukula:1994mn} in which a lower boson
mass limit of 375 GeV is claimed.

There are similar processes for the tau lepton:  $\tau \to \mu\mu\mu$,
$\tau\to ee\mu$, $\tau \to \mu\mu e$, and $\tau \to eee$.  However,
the branching ratios for all of these are on the order of ${\cal
O}(10^{-6})$ and place too weak of constraints to be of relevance in
this study \cite{Groom:2000in}.

\subsection{$\mu \to e \gamma$}

$\mu^+ \to e^+ \gamma$ has historically been probed by allowing an antimuon at
rest to decay and waiting for back-to-back decay products.  For the
same reasons as $\mu \to eee$, a negative muon is not used.  The
dominant diagrams for $\mu \to e\gamma$ are identical to the muon
$g-2$ with the outgoing antimuon exchanged for a positron.  This
provides a new probe by changing the charges probed at the vertices
involving the positron.  In principle, $\mu \to e\gamma$ probes all
charges of model $X$.  However, a tau internal propagator dominates for
the same reasons it should dominate lepton anomalous magnetic
moments.  This effectively reduces the charges probed to only $Q^l_{13}$
and $Q^l_{23}$.

The partial width for $\mu \to e\gamma$ is

\begin{equation}
\Gamma(\mu \to e\gamma) = \frac{\alpha G_{\rm F}^2}{4\pi^4}
\left( \frac{g_X}{g_Y} \right)^4
\frac{m_W^4 m_\mu^3 M_{12}^2}{m_X^4}
\end{equation}

\noindent
where
\begin{equation}
M_{12} = m_e Q^l_{11} Q^l_{12} + m_\mu Q^l_{12} Q^l_{22}
+ m_\tau Q^l_{13} Q^l_{23}.
\end{equation}

\noindent
For model $X$, the branching ratio results in

\begin{equation}
R(\mu \to e\gamma) = 1.3\times10^{-13}
\left( \frac{g_X}{g_Y} \right)^4
\left( \frac{Q^l_{13} Q^l_{23}}{10^{-5}} \right)^2
\left( \frac{1\,{\rm TeV}}{m_X} \right )^4.
\end{equation}

\begin{figure}[!btp]
\center \resizebox{4 in}{!}{\includegraphics{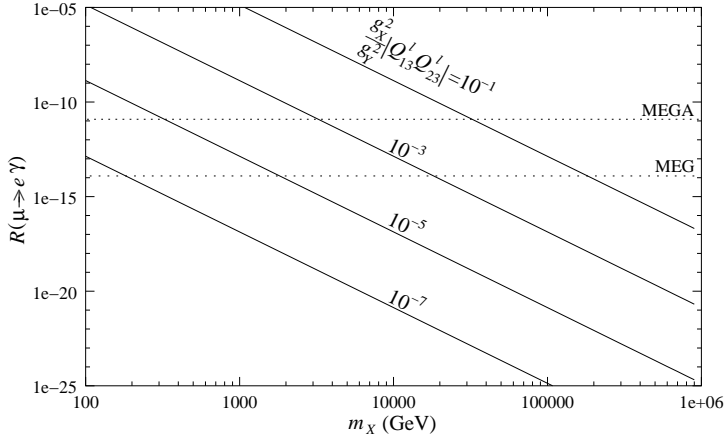}}
\caption{LFV boson mass $m_X$ (GeV) vs.\ the branching ratio $R(\mu
\to e\gamma)$.  It is important to eliminate the possible confusion
over which experiments are most constraining for a given model.  The value
of this plot is in revealing that the $\mu \to e\gamma$ MEGA limit is of
little value to constrain model $X$ bosons beyond or
complementary to muon conversion when BNL E821 muon $g-2$ constraints
are applied.  The muon $g-2$ optimal charges are $|Q^l_{13}
Q^l_{23}|=10^{-5}$ and such, MEGA has probed model $X$ bosons to about
200 GeV for $g_X=g_Y$.  If muon $g-2$ constraints are relaxed, with
large $e$-$\tau$ charge $Q^l_{13}$, the MEGA limit has probed to
very large LFV $Z'$ boson masses, i.e. 10 TeV for $(g_X/g_Y)^2
|Q^l_{13} Q^l_{23}| = 10^{-2}$.  This plot does not assume the LFV should
account for all of the muon $g-2$ observed deviation.}
\label{fig.mueg}
\end{figure}

\noindent
The current limit is held by MEGA at LANL (1999) to be
$R(\mu \to e\gamma) < 1.2\times10^{-11}$ \cite{Brooks:1999pu}.  A
recently approved experiment MEG at PSI may reach a sensitivity of
$10^{-14}$ \cite{psi} in 2003.  From Fig.\,\ref{fig.mueg},
it is seen that the muon conversion limit overrides the $\mu \to
e\gamma$ limit when considering the parameter space covered in model $X$.
For example, using the muon $g-2$ optimal charges again,
$|Q^l_{13}Q^l_{23}|=1$, $\mu \to e\gamma$ has only probed the model $X$
boson up to about 200 GeV.  An earlier effort confirmed this ranking
of LFV experiments for top-color assisted technicolor, a specific
case of model $X$ \cite{Rador:1999is}.

Due to the relatively weak branching ratio
limits for $\tau \to \mu\gamma$ and $\tau \to e\gamma$ (both on the
order of $10^{-6}$, those analyses are rendered irrelevant for model
$X$ \cite{Groom:2000in}.  However, a stronger limit on $\tau \to
\mu\gamma$ could prove interesting since the charges involved,
$Q^l_{23}$ and $Q^l_{33}$, may have magnitudes near 1 for the muon
$g-2$ optimal charges.

\subsection{$e^+e^- \to \mu^+\mu^-$}

$e^+e^- \to l^+l^-$ is included in our analysis to provide a simple
way of constraining the mass of any $Z'$ model at future linear
colliders.  Because the cross-section $\sigma(e^+e^- \to l^+l^-)$ is
insensitive to what outgoing charged lepton is used, we arbitrarily
choose muons as outgoing.  At tree level, this process is dominated by the
$s$-channel exchange involving only charges $Q^l_{ii}$.  $t$-channel
exchange suffers $(Q^l_{12})^4$ suppression, rendering it relevant only
for $X$ bosons with masses somewhere over 1,000 TeV where the
$e$-$\mu$ coupling $(g_X/g_Y)^2 Q^l_{12}$ may be 1 or larger
(Fig.\,\ref{fig.muconv}).  This would come at the expense of losing a
LFV $Z'$ interpretation of the muon $g-2$ observed discrepancy.

\begin{figure}[!btp]
\center \resizebox{4 in}{!}{\includegraphics{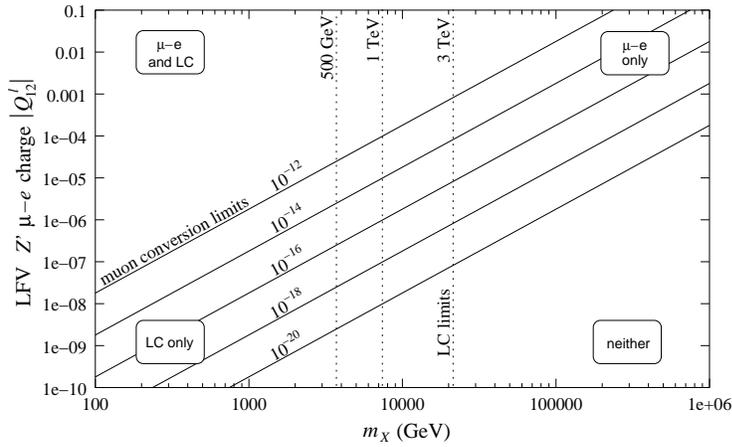}}
\caption{LFV $Z'$ boson mass $m_X$ (GeV) vs.\ the electron-muon charge
$Q^l_{12}$.  This plot demonstrates future muon conversion and linear
collider experiments to be highly complementary.  The vertical lines
show the maximum $Z'$ mass that would show a 1\% deviation in
$\Delta\sigma(e^+e^- \to \mu^+\mu^-)/\sigma_{\rm SM}(e^+e^- \to
\mu^+\mu^-)$ with parameters $g_X=g_Y$, $Q^l_{11}=-1$, and
$Q^l_{22}=-1$.  The diagonal lines are limits that would be set by
future muon conversion experiments.  The upper left ``quadrant''
contains parameter space that would be implied at both types of
experiments.  The upper right quadrant contains parameters that could
only be implied at future muon conversion experiments.  The lower left
quadrant contains parameters that could only be implied by future
linear colliders.  Neither type of experiment could see the parameters
of the lower right quadrant.}
\label{fig.nlc}
\end{figure}

The next linear colliders have potential to show a 1\% deviation in
the cross-section for $e^+e^- \to l^+l^-$. It is assumed a 1\%
observed difference in $\Delta\sigma(e^+e^- \to 
\mu^+\mu^-)/\sigma_{\rm SM}(e^+e^- \to \mu^+\mu^-)$ is entirely
due to the $X$ boson.  We include interference with photon and $Z$
exchange in $\Delta\sigma$.  Fig.\,\ref{fig.nlc} has the parameter space
limits set forth by future linear colliders superimposed with the
limits set by muon conversion.  Parameters $g_X=g_Y$, $Q^l_{11}=-1$,
and $Q^l_{22}=-1$ are used.

It is seen that the future muon
conversion and linear collider experiments are highly complementary in
their search for a model $X$ boson.  Signals implying parameters in
the upper left ``quadrant'' of $m_X$ and $Q^l_{12}$ parameter space
can be seen at both types of experiments.  The larger masses in the
upper left quadrant can be seen only by muon conversion experiments.
Due to the insensitivity to $t$-channel exchange, future linear
colliders may probe the region of arbitrarily small $e$-$\mu$
charge in the lower left corner.  Neither type of experiments will see
the lower right corner.

\section{Summary}
Theoretical and experimental motivations for a lepton-flavor violating
$Z'$ boson are explored.   A LFV $Z'$ boson interpretation is applied
to recent and near-future experiments.  A conservative model was chosen by
balancing the least number of free parameters while not sacrificing
generic phenomenological behavior.  Through this ``model $X$,'' we
established the following:

\begin{itemize}
\item The BNL E821 muon $g-2$ deviation may fully be attributed to a
$X$ boson with a mass as large as roughly 6 TeV, or less as the
coupling constant $g_X$ is lowered.  This LFV boson will also maintain
compatibility with muon conversion, $\mu \to eee$, and $\mu\to
e\gamma$ experimental null searches.

\item Even without LFV, a $Z'$ with only lepton-flavor conserving
interactions may still account for the muon $g-2$ deviation with
masses as large as 660 GeV.  No linear collider may place a limit on this
mass due to the parameter space that allows for arbitrarily small
$eeX$ couplings.

\item Model $X$ is immediately testable.  Nearly all parameter space
points in the first claim above will have produced one to twenty
$e^+e^- \to \mu\tau$ events at LEP II.  An analysis of the LEP II data
is therefore urged.

\item The technological advances incorporated in future muon
conversion experiments, MECO (2006) and PRIME (2007), will improve
probes of LFV $Z'$ masses by more than an order of magnitude.  For
example, 10 TeV bosons with couplings of $10^{-5}$ will be accessible.

\item 
Furthermore, muon conversion places the strictest lower bounds on LFV
$Z'$ masses over all other experiments, including $\mu\to eee$,
$\mu\to e\gamma$, and all collider searches.  MEG, a future $\mu \to
e\gamma$ experiment, will not probe model $X$ bosons beyond what muon
conversion already has.  With no forthcoming $\mu \to eee$ experiments
announced, the  importance of MECO and PRIME for the next decade is
emphasized.

\item Future linear colliders will be complementary in their search
for $Z'$ bosons.  As they are insensitive to LFV primal vertices,
linear colliders provide different limits as explained in
Fig.\,\ref{fig.nlc}.

\end{itemize}

\noindent
{\bf Acknowledgements.}
We thank S.~Mrenna, K.~Tobe, and J.~Wells for helpful discussions.
This work was funded in part by the Department of Energy.


\end{document}